\documentclass[sigconf,screen]{acmart}
\usepackage{multirow}
\usepackage{float}            % [H] for appendix floats
\usepackage{pifont}
\usepackage{array}
\usepackage{xspace}
\usepackage{algorithm}
\usepackage[noend]{algpseudocode}
\usepackage{tikz}
\usepackage{tabularx}
\usepackage{xcolor}
\definecolor{findingbg}{HTML}{EEF3F6}

\usepackage{colortbl}       % \rowcolor for table header rows
\usetikzlibrary{arrows.meta, positioning, fit, backgrounds, calc}
\usepackage{enumitem}
\setlist{leftmargin=*, itemsep=1pt, topsep=2pt}
\AtBeginDocument{%
  }

\newcommand{\sys}{\textit{Janus}\xspace}
\newcommand{\TODO}[1]{\textcolor{red}{\relax\ifmmode\text{#1}\else#1\fi}}
\newcommand{\tabhead}{\rowcolor{gray!15}}
\newcommand{\tabheadfirst}[2]{\multicolumn{1}{@{}>{\columncolor{gray!15}[0pt][\tabcolsep]}#1}{#2}}
\newcommand{\tabheadlast}[2]{\multicolumn{1}{>{\columncolor{gray!15}[\tabcolsep][0pt]}#1@{}}{#2}}
\newcommand{\tabsetup}{\setlength{\aboverulesep}{0pt}\setlength{\belowrulesep}{0pt}\renewcommand{\arraystretch}{1.25}}
\newcommand{\para}[1]{\smallskip\noindent\textbf{#1}}
\newcounter{findingctr}
\newcommand{\finding}[1]{%
  \refstepcounter{findingctr}%
  \smallskip\noindent
  \fcolorbox{black}{findingbg}{%
    \parbox{\dimexpr\linewidth-2\fboxsep-2\fboxrule\relax}{%
      \small\textbf{Finding \thefindingctr.} #1%
    }%
  }%
  \smallskip
}
\newcommand{\Wl}{\textsf{G}}            % gateway (WAF) layer
\newcommand{\Al}{\textsf{A}}            % application layer
\newcommand{\Sl}{\textsf{S}}            % system runtime layer

\newcommand{\ASR}{\mathrm{ASR}}

\newcommand{\BPR}{\mathrm{BPR}}
\newcommand{\Atk}{\mathcal{A}}
\newcommand{\Ben}{\mathcal{B}}

\setcopyright{acmlicensed}
\copyrightyear{2027}
\acmYear{2027}
\acmDOI{XXXXXXX.XXXXXXX}
\acmISBN{978-1-4503-XXXX-X/2027/XX}

\begin{document}

\title{JANUS: Synthesizing Cross-Layer Defenses Against Web Deserialization Attacks}

\author{Ruozhao Yang}
\email{rz.yang.2024@phdcs.smu.edu.sg}
\orcid{0009-0004-2893-5171}
\affiliation{%
  \institution{Singapore Management University}
  \country{Singapore}
}

\author{Minglang Li}
\email{2024060909004@std.uestc.edu.cn}
\orcid{0009-0007-4209-8877}
\affiliation{%
  \institution{UESTC}
  \country{Chengdu, China}
}

\author{Jiongchi Yu}
\email{jiongchiyu@acm.org}
\orcid{0000-0002-2888-4499}
\affiliation{%
  \institution{Nanyang Technological University}
  \country{Singapore}
}

\author{Xinfeng Li}
\email{xinfeng.li@polyu.edu.hk}
\orcid{0000-0002-9686-4369}
\affiliation{%
  \institution{The Hong Kong Polytechnic University}
  \country{Hong Kong, China}
}

\author{Xiaofei Xie}
\email{xfxie@smu.edu.sg}
\orcid{0000-0002-1288-6502}
\affiliation{%
  \institution{Singapore Management University}
  \country{Singapore}
}

\renewcommand{\shortauthors}{Yang et al.}

\begin{abstract}
Unsafe deserialization turns untrusted data into attacker-controlled execution, and a single vulnerability can admit diverse exploit variants that evade defenses tailored to a known exploit. Existing defenses typically operate at a single enforcement point with limited visibility into the full attack. Combining controls across layers offers complementary protection, but coordinating them to preserve legitimate functionality and adapting them to new variants requires substantial expert effort.

We propose \sys, a red-blue adversarial framework for automated cross-layer defense synthesis. Starting from a known exploit, \sys iteratively refines deployment-specific policies in an isolated offline environment. A black-box attacker probes the current policy for exploit variants using only externally observable feedback. The defender summarizes each verified variant into its security-relevant effects, uses counterfactual replay to separate complementary coverage from redundant enforcement, and jointly selects and scopes controls across layers. Each candidate policy is validated against accumulated exploits and benign workflows.

Across sixteen vulnerability deployments in six Java and PHP applications, \sys reduces held-out attack success to 6.2\% versus 18.8-55.0\% for gateway, application, system, and cross-layer baselines. It fully mitigates 13 deployments and passes 98.4\% of benign functionality tests, with a 6.9\% increase in targeted P95 latency.

% We evaluate \sys on sixteen vulnerability deployments across six Java and PHP
% applications against gateway, application, system, and cross-layer defenses.
% \sys reduces held-out attack success to 6.2\%, compared with 18.8-55.0\% for the
% baselines, fully mitigates 13 of 16 deployments, and preserves 98.4\% of benign
% functionality tests at a 6.9\% targeted P95 latency increase.

\end{abstract}




\maketitle

\section{Introduction}
\label{sec:intro}

Unsafe deserialization poses a persistent and critical threat to Web security~\cite{owasptop10,sayar2023deserstudy}. When applications reconstruct objects from untrusted data, attacker-controlled types and field values can trigger chains of existing methods that lead to remote code execution (RCE)~\cite{sayar2023deserstudy,peng2026kintsugi}. Existing defenses constrain specific steps of this process, for example by restricting the classes permitted during deserialization~\cite{zhang2024deseriguard,david2024quack}. However, a single vulnerability may admit multiple gadget chains drawn from application dependencies, so filters targeting known gadgets can be bypassed by alternative chains~\cite{cao2023oddfuzz,chen2023tabby,park2022fugio,sayar2023deserstudy}. Mitigation is also deployment-specific: the deployed libraries and runtime determine which chains are viable~\cite{kreyssig2024prerequisites}, whereas legitimate features determine which deserialization behaviors must remain available~\cite{huang2016talos}.

\begin{figure}[t]
  \centering
  \includegraphics[width=0.75\columnwidth]{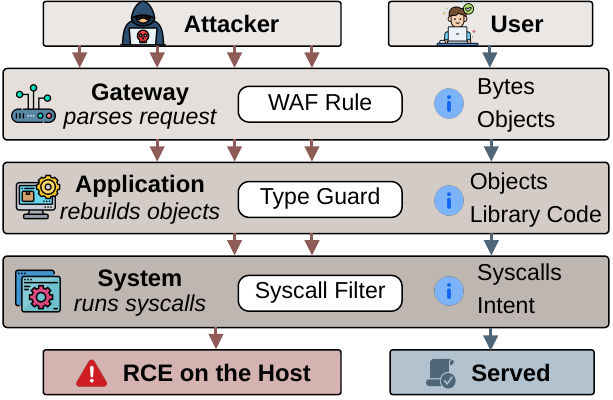}
  \caption{Motivating example. Each layer's defense stops some exploit variants, but one variant passes all three layers and reaches RCE, while the legitimate request is served.}
  \Description{An attacker and a user send requests down through gateway, application,
  and system bands. Three attack attempts stop at different bands, one reaches RCE, and
  the user's request is served.}
  \label{fig:motivation}
\end{figure}

\textbf{Motivation.} As shown in~\autoref{fig:motivation}, defenders can intervene at three layers with distinct visibility and established enforcement mechanisms. At the gateway, a Web application firewall (WAF) inspects and filters incoming request representations~\cite{owaspcrs}. Application-level deserialization filters constrain the types and object structures accepted during reconstruction~\cite{oracleSerializationFiltering,zhang2024deseriguard,david2024quack}. System-level policies restrict the resulting operations through system-call filtering or container isolation~\cite{bulekov2021saphire,lei2025rightcalls,peng2026kintsugi}. Each layer nevertheless observes only part of the attack: narrow restrictions may miss exploit variants, while broader restrictions risk disrupting legitimate features~\cite{sayar2023deserstudy,demetrio2020wafamole}.

Combining these layers offers complementary protection~\cite{rubel2005defenseindepth}, but requires coordinating their policies. We define a \textit{multi-layer composition} as a joint selection of controls with deployment-specific parameters and enforcement scopes. Constructing an effective composition is challenging because upstream blocking can obscure downstream coverage, overlapping controls may introduce redundant overhead, and policy interactions can disrupt legitimate workflows in ways that isolated checks do not reveal. Moreover, a composition that blocks known exploits may still admit variants using different gadget chains or payload representations. Existing automated mitigations support the generation of security workarounds, deserialization restrictions, or runtime policies~\cite{huang2016talos,zhang2024deseriguard,peng2026kintsugi}, but these capabilities alone do not establish how controls should be jointly configured and revised to cover adaptive exploit variants while preserving legitimate functionality.

\textbf{Our Approach.} We present \sys, a red-blue adversarial framework for synthesizing deployment-specific multi-layer defenses against unsafe deserialization. The framework pairs an attacker agent with a defender agent on an isolated replica of the deployment. Starting from a known exploit, the attacker adapts its gadget-chain selection and payload representation to achieve a specified consequence, such as command execution within the application container. The defender uses execution evidence and benign-workflow feedback to jointly revise controls across layers. Through this offline interaction, the agents expose and address weaknesses beyond those revealed by the initial exploit. The attacker observes only what an external attacker could observe and serves as a source of counterexamples. The technical contributions of \sys lie on the defense side, in how the defender interprets these counterexamples and coordinates controls to cover them.

Realizing this process raises three technical challenges. \ding{182}~\textbf{Exposing Variants Under Black-Box Observation.} A defense is only as robust as the variants it is checked against, yet an external attacker cannot distinguish a gateway rejection from an incompatible gadget chain. \sys's attacker parameterizes reviewed fixtures whose prerequisites match the deployment, treats ambiguous feedback as hypotheses, and submits candidates to an independent oracle that it cannot observe. \ding{183}~\textbf{Interpreting Exploit Effects and Defensive Coverage.} Raw system-call traces are too low-level for policy reasoning, and upstream blocking hides the protection of downstream controls. \sys condenses execution evidence into attack-effect summaries expressed as security-relevant capabilities, each tied to the control that can intercept it, if any, and uses counterfactual replay with upstream rules disabled to separate complementary coverage from redundant enforcement. \ding{184}~\textbf{Coordinating Controls Under Functional Constraints.} Controls that are individually acceptable may break legitimate workflows when combined, and some improvements require coordinated changes across layers. \sys jointly selects and scopes controls from a bounded catalog of deployable mechanisms and accepts a candidate only if it blocks all verified exploits and meets the functionality and latency thresholds on the benign-workflow regression suite.

% Realizing this loop requires addressing three technical challenges.
% (1)~\emph{Perceiving the exploitation state.} Under black-box observation, a
% request rejected at the gateway and a payload whose gadget never fired return
% similar responses, so the attacker cannot tell what to change. \sys localizes each
% failure to an exploitation stage from runtime evidence and an independent
% consequence oracle, and drives exploration with a stage-based state machine that
% decides whether to switch chains or only mutate the representation.
% (2)~\emph{Understanding the attack's effect.} Raw system-call traces are too large
% and low-level for the defender to reason about, and once an upstream rule rejects
% an exploit, the coverage of downstream rules becomes unobservable. \sys condenses
% causally aligned evidence into a compact attack-effect summary expressed in the
% capabilities that system-layer controls govern, and replays exploits with upstream
% rules masked to attribute coverage to each layer.
% (3)~\emph{Composing layers without breaking functionality.} Per-layer rules
% interact and may harm legitimate workflows, and an effective fix sometimes exists
% only as a pair of edits that each fail alone. \sys has the defender propose
% controls per layer from a bounded catalog of deployable mechanisms and compose them
% into one policy, and accepts the policy only if it blocks every past exploit and
% passes the benign workflows. Together, these designs close the loop into a
% self-evolving process in which each round of attack sharpens the next round of
% defense.

\textbf{Evaluation.} We compare \sys on 16 deployments across six applications
against ModSecurity with OWASP CRS~\cite{owaspcrs}, Datadog RASP and
OpenRASP~\cite{datadograsp,openrasp}, Kintsugi~\cite{peng2026kintsugi}, and
Tetragon~\cite{tetragon}. \sys lowers held-out attack success to 6.2\%, versus
18.8-55.0\% for the baselines, fully mitigates 13 of 16 deployments, and is
breached on 5 of 48 unseen-attacker/deployment pairs, while passing 98.4\% of
functionality tests. A recorded run on CVE-2026-84645, a recent Jenkins RCE,
illustrates the refinement loop.

\textbf{Contributions.} This paper makes three contributions:
\begin{itemize}
\item We formulate unsafe-deserialization mitigation as deployment-specific cross-layer defense synthesis that coordinates complementary controls under functional constraints.
\item We propose \sys, a red-blue adversarial framework in which a black-box attacker supplies verified exploit variants and a defender combines attack-effect summaries with counterfactual coverage attribution to jointly select and scope cross-layer controls under functional constraints.
\item We define an evaluation protocol with held-out paths, unseen attackers, multi-consequence oracles, and benign-workflow checks, and evaluate \sys under this protocol.
% We evaluate \sys under this protocol and will release its implementation and evaluation materials upon publication\footnote{Anonymous repository: \url{https://anonymous.4open.science/r/VELA-70DF/}}.
\end{itemize}
\section{Background and Problem Statement}
\label{sec:problem}
\label{sec:background}

\subsection{Deserialization Attacks and Defenses}
\label{sec:bg}

Unsafe deserialization allows attacker-controlled data to trigger unintended method invocations during object reconstruction. A sequence of existing methods leading to a sensitive operation constitutes a gadget chain~\cite{dahse2014pop,sayar2023deserstudy}. A vulnerable entry point may admit multiple chains~\cite{park2022fugio,cao2023oddfuzz,chen2023tabby}, whose viability depends on the deployed libraries and runtime~\cite{kreyssig2024prerequisites,cao2023overriding}. Exploit variants can differ in where serialized data enters, how it is encoded and framed, or which execution path leads to the consequence.

Each of the three defensive layers has a characteristic blind spot. Gateway controls (\Wl) such as reverse proxies and WAFs~\cite{owaspcrs} inspect request representations rather than reconstructed objects, so representation changes can bypass them~\cite{wu2025wafbooster,demetrio2020wafamole}. Application controls (\Al) restrict accepted classes or object-graph properties where the parser exposes them~\cite{jep290,oracleSerializationFiltering}, yet may still permit malicious uses of allowed classes. System controls (\Sl) restrict process creation, file access, or network communication~\cite{dockerSeccomp,bulekov2021saphire}, limiting consequences without preventing deserialization, often for a whole process or container. These controls also differ in availability across deployments, so an effective composition must account for both the controls a deployment supports and the legitimate behavior within their scopes.

\subsection{Threat Model}
\label{sec:problem:threat}

We consider a Web deployment $D$ with a known, unpatched deserialization vulnerability and a validated seed exploit that achieves a specified security consequence. The defender operates on isolated, resettable replicas with representative benign workflows.

\textbf{Attacker.} The attacker controls HTTP requests and holds the privileges required to reach the vulnerable entry point. It may adapt payload representations and gadget chains using responses, timing, and attacker-controlled out-of-band observations within fixed query and time budgets, but cannot alter deployment configurations, defensive policies, or validation infrastructure. The offline attacker agent operates under the same restrictions and receives no internal execution-stage feedback.

\textbf{Defender.} The defender analyzes application logs, network traffic, and system logs to configure supported controls across the three layers. It may change only configuration, not application source code or business logic.

\textbf{Trust Assumptions.} The host kernel, enforcement backends, and validation oracles are trusted. An independent oracle verifies exploit success through controlled state or effects rather than responses or logs alone.

\textbf{Security Objective.} A defense must prevent the specified consequence, and intervention after it occurs does not count. We distinguish blocking before or during deserialization from impact limitation that denies a subsequent operation. We assess protection only for the specified consequences, attackers, and budgets.

\subsection{Cross-Layer Defense Synthesis}
\label{sec:problem:def}

Given a deployment $D$ and a validated seed exploit, our goal is a multi-layer defense against the original attack and its variants that exploit the same vulnerability through alternative representations or execution paths. Let $\Pi_D$ denote the configurations supported by the deployment's controls and scopes, and $\Ben$ the benign workflows that pass without the defense. Refinement alternates between defense configuration and bypass exploration, and verified attacks accumulate in an archive $\Atk_t$ initialized with the seed. At iteration $t$, the defender seeks a configuration
{\setlength{\abovedisplayskip}{3pt}\setlength{\belowdisplayskip}{3pt}%
\begin{equation}
\label{eq:objective}
\pi\in\Pi_D\ \ \text{s.t.}\ \ \ASR(\pi,\Atk_t)=0,\ \BPR(\pi,\Ben)\ge\tau_b,\ \Delta L_{95}(\pi)\le\epsilon_l,
\end{equation}}%
where $\ASR$ is the fraction of attacks achieving their specified consequences, $\BPR$ the fraction of workflows passing all functional assertions, and $\Delta L_{95}$ the P95 latency increase of the targeted endpoint under benign traffic, with thresholds $\tau_b$ and $\epsilon_l$ fixed before refinement begins.

Because every archived attack is a verified exploit, the archive must be covered completely rather than traded off against functionality, so Eq.~\eqref{eq:objective} is a feasibility problem. New bypasses expand the archive and trigger further revision, and failure to meet the conditions within budget is recorded as unsuccessful. The final configuration is frozen before independent evaluation against attacks excluded from refinement.

\section{Design of \sys}
\label{sec:design}

\begin{figure*}[t]
  \centering
  \includegraphics[width=\textwidth]{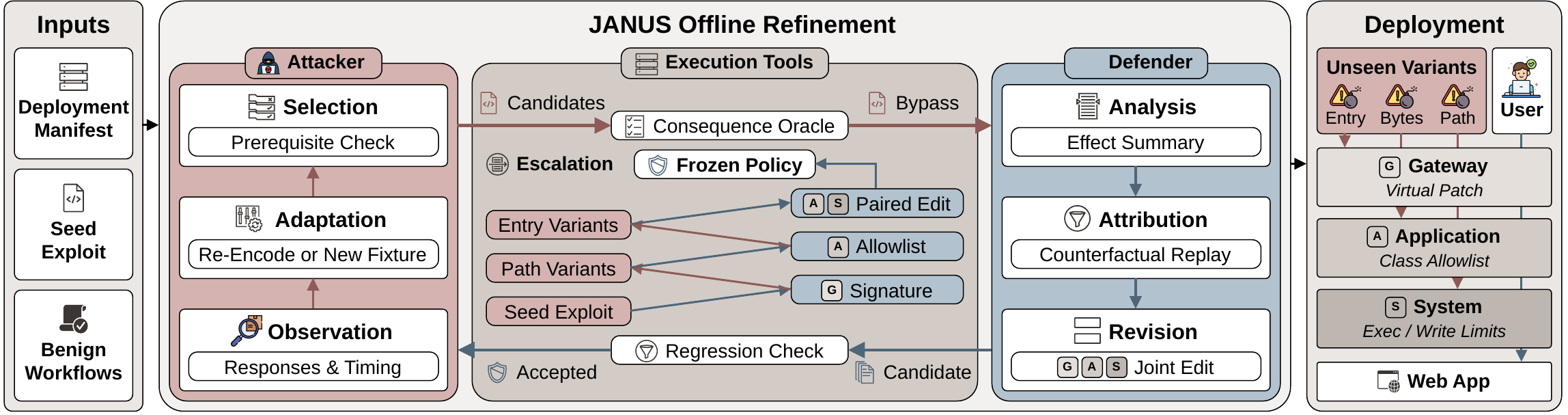}
  \caption{Overview of \sys, which alternates black-box attack exploration and evidence-guided defense refinement under regression checks.}
  \Description{An attacker explores bypasses against a fixed configuration. A defender analyzes attack evidence, proposes controls, and refines them using benign-workflow feedback. Regression checks determine whether an update is accepted or returned for revision.}
  \label{fig:arch}
\end{figure*}

\sys searches for a configuration satisfying Eq.~\eqref{eq:objective} by alternating two roles on isolated replicas of $D$ (Figure~\ref{fig:arch}). Starting from the archive $\Atk_0=\{x_0\}$, the defender turns attack evidence and benign-workflow feedback into a configuration $\pi\in\Pi_D$, which is accepted only if it blocks every archived attack and meets both thresholds. The attacker then probes the accepted configuration with externally observable feedback, and each verified bypass enlarges the archive and triggers another revision. The last accepted configuration is frozen and deployed without language-model decisions.

\subsection{Black-Box Attack Exploration}
\label{sec:design:attacker}

The attacker supplies counterexamples and observes only what an external attacker could. This is hard because exploits depend on deployment-specific classes and runtime behavior~\cite{kreyssig2024prerequisites,cao2023overriding}, and a rejected request, an incompatible gadget, and a denied downstream operation can look identical from outside.

\textbf{Fixture Catalog.}
The attacker draws from a bounded catalog of reviewed fixtures, each specifying its prerequisites, entry point, payload construction, and permitted variations of entry point, representation, or execution path. A fixture is eligible unless known target information contradicts its prerequisites. Held-out fixtures are disjoint from the catalog by path key (Appendix~\ref{app:referee}) and stay hidden during refinement, and a permitted variation that reaches a held-out path counts as rediscovery.

\textbf{Observation-Guided Exploration.}
In each round the configuration $\pi$ is fixed, and the attacker returns candidates $X$ generated from eligible fixtures. It sees only the response, timing, and attacker-controlled out-of-band signals of its own requests, never logs or policy text, and treats them as hypotheses rather than execution-stage labels~\cite{pham2020aflnet}. Apparent filtering suggests a representation change, an exposed compatibility error suggests another fixture, and an ambiguous failure leaves both open.

\textbf{Counterexample Verification.}
An independent oracle $\omega(x,\pi)\in\{1,0,\bot\}$ reports whether attack $x$ reaches its specified consequence under $\pi$, fails to reach it, or cannot be decided. A candidate becomes a counterexample only if it succeeds both without and with the defense:
\begin{equation}
\label{eq:verify}
V(X,\pi)=\{x\in X:\ \omega(x,\emptyset)=1\ \wedge\ \omega(x,\pi)=1\}.
\end{equation}
Failure on the undefended replica is thus never credited to the defense, $\bot$ never counts as prevention, and $\omega$ is never returned to the attacker.

\subsection{Evidence-Guided Defense Refinement}
\label{sec:design:defender}

The defender maps execution evidence onto deployable controls and reconciles protection with legitimate use.

\textbf{Attack-Effect Summary.}
For each archived attack, \sys links the request to application and system events through request identifiers and process lineage. The summary records the oracle-confirmed consequence, the exercised capabilities with their decisive parameters (Table~\ref{tab:aes}), and the enforcement events under the current configuration, while unlinked or missing observations remain unknown. It ties behavior to intervention points without requiring the defender to reconstruct the gadget chain.

\textbf{Rules.}
A rule $r=(c,\sigma,\theta)$ instantiates a catalog control $c$ (Table~\ref{tab:catalog-main}) with scope $\sigma$ and parameters $\theta$, and carries its supporting evidence and expected effect. A configuration $\pi$ is a set of rules, and $\Pi_D$ admits only controls and scopes that $D$ supports.

\textbf{Counterfactual Coverage Attribution.}
Once an upstream rule rejects $x$, downstream protection is unobserved but may exist. \sys replays $x$ from the same snapshot, masks the rule that blocked it, and repeats until $x$ reaches its consequence or no further rule blocks it (Appendix~\ref{app:referee}). The result is a coverage matrix
\begin{equation}
\label{eq:cov}
\mathit{Cov}[r][x]\in\{\mathrm{obs},\mathrm{lat},\mathrm{none},\mathrm{unk}\},\qquad r\in\pi,\ x\in\Atk,
\end{equation}
where obs marks the first rule that blocks $x$, lat a rule that blocks it only once upstream rules are masked, none a rule still enabled when the consequence occurs, and unk an inconclusive replay. Let $C(x)$ be the rules with obs or lat for $x$. On rows without unk, $x$ has a \emph{single point of failure} when $|C(x)|=1$, and a rule is a \emph{redundancy candidate} when every attack it covers has $|C(x)|\ge 2$. Because this test is pairwise, a removal is confirmed only by the acceptance check on the complete configuration. Replaying the training workflows in the same way records which workflows each rule breaks.

\textbf{Joint Revision.}
From the summaries, the coverage matrix, and failed workflow assertions, the defender proposes a revision that may add, remove, narrow, or retune rules at several layers at once. Single points of failure are preserved, redundancy candidates may be removed or narrowed, and a rule that breaks a workflow is narrowed or replaced by a control at another layer covering the same consequence. Individual edits need not be acceptable alone, since only the complete revised configuration is checked.

\subsection{Regression-Guarded Evolution}
\label{sec:design:evolution}

\textbf{Acceptance.}
With training workflows $\Ben_{\mathrm{tr}}\subseteq\Ben$ and the thresholds of Eq.~\eqref{eq:objective}, a candidate $\pi$ is accepted if and only if
\begin{equation}
\label{eq:acc}
\begin{aligned}
\mathrm{Acc}(\pi,\Atk)\;\equiv\;{}&\textstyle\bigwedge_{x\in\Atk}\,\omega(x,\pi)=0\\
&\wedge\ \BPR(\pi,\Ben_{\mathrm{tr}})\ge\tau_b\ \wedge\ \Delta L_{95}(\pi)\le\epsilon_l
\end{aligned}
\end{equation}
on a replica where the native configuration has loaded, and any $\bot$ falsifies the predicate. Rejected candidates return to the defender with their failure evidence but never replace the incumbent.

\textbf{Rounds.}
A round is one attack phase against the fixed incumbent $\pi_t$. It ends when the attacker returns $X_t$ or exhausts its per-round query allowance. The archive then grows to $\Atk_{t+1}=\Atk_t\cup V(X_t,\pi_t)$, and only a new bypass triggers a revision, so the attacker never interacts with a partially updated candidate (Algorithm~\ref{alg}). Exploration, revision, and checking all consume the budget $Q$.

\begin{algorithm}[t]
\caption{Workflow of \sys}
\label{alg}
\small
\begin{algorithmic}[1]
\Require Deployment $D$, seed $x_0$, workflows $\Ben_{\mathrm{tr}}$,
         budget $Q$, threshold $k$
\Ensure Frozen configuration $\pi^\star$ or \textit{RefinementFailure}
\State $\pi \gets \emptyset$, $\Atk \gets \{x_0\}$,
       $E \gets \textit{Evidence}(x_0)$
\State $u \gets \textbf{true}$, $n \gets 0$
\While{$\textit{BudgetRemains}(Q)$ \textbf{and} $n < k$}
    \If{$u$}
        \State $(S,C) \gets
            \textit{Analyze}(\pi,\Atk,\Ben_{\mathrm{tr}})$
            \Comment{summaries and coverage}
        \State $\pi' \gets
            \textit{JointRevise}(\pi,S,C,E,\Ben_{\mathrm{tr}})$
        \State $(a,F) \gets
            \textit{Check}(\pi',\Atk,\Ben_{\mathrm{tr}})$
        \State $E \gets E \cup F$
        \State \textbf{if not} $a$ \textbf{then continue}
        \State $\pi \gets \pi'$, $u \gets \textbf{false}$
    \EndIf
    \State \textbf{if not} $\textit{BudgetRemains}(Q)$
           \textbf{then break}
    \State $(X,e) \gets \textit{AttackRound}(D[\pi])$
        \Comment{$e$: catalog exhausted}
    \State $Y \gets V(X,\pi)$, $n \gets n+1$
    \If{$Y \neq \emptyset$}
        \State $\Atk \gets \Atk \cup Y$,
               $E \gets E \cup \textit{Evidence}(Y)$
        \State $u \gets \textbf{true}$, $n \gets 0$
    \ElsIf{$e$}
        \State \textbf{break}
    \EndIf
\EndWhile
\State \Return $\textit{Finalize}(\pi,\Atk,\Ben_{\mathrm{tr}})$
\end{algorithmic}
\end{algorithm}

\textbf{Stopping and Freezing.}
Refinement stops at \emph{budget exhaustion}, at \emph{stagnation} after $k$ consecutive rounds without a new verified bypass, or at \emph{catalog exhaustion}, when every eligible fixture and permitted variation has been tried against the incumbent. None of them proves that no bypass exists. \textit{Finalize} re-checks the last incumbent on a fresh replica and returns it as $\pi^\star$ if it is still accepted, or \textit{RefinementFailure} otherwise. Held-out attacks and test workflows are used only to evaluate $\pi^\star$.

\textbf{Run-to-Run Variation.}
Both agents sample from an LLM, so runs from the same seed may return different configurations, whereas verification, replay, and regression checks are deterministic given the snapshot. Every returned configuration blocks its own archive and passes the training workflows, but its coverage beyond that archive varies, so we report the spread over repeated runs.

\section{Implementation}
\label{sec:impl}

We implement \sys in about 9,800 lines of Python. A control plane runs Algorithm~\ref{alg}, both agents share one LLM backend with schema-validated outputs (Appendix~\ref{app:prompts}), and verification, counterfactual replay, and regression checks are deterministic tools (Appendix~\ref{app:referee}). Each deployment runs as a Docker Compose project behind an NGINX reverse proxy and is restored from a snapshot before every replay. Oracles and workflow assertions run outside the application container, and failed replays are retried rather than counted as blocks. No application code is modified.

\begin{table}[t]
\centering
\caption{Defensive controls supported by \sys at each layer. B: attack blocking. L: limitation of a specified effect.}
\label{tab:catalog-main}
\footnotesize\setlength{\tabcolsep}{3pt}\tabsetup
\begin{tabularx}{\columnwidth}{@{}l|l >{\raggedright\arraybackslash}X c@{}}
\toprule
\tabhead \tabheadfirst{l|}{ID} & Control & Configurable Policy & \tabheadlast{c}{Effect} \\
\midrule
G1 & Ingress Limits & Allowed routes and methods, body-size cap & B \\
G2 & Virtual Patch & Request-match rules and action & B \\
\midrule
A1 & Class Filtering & Class allowlist, others rejected & B \\
A2 & Graph Limits & Depth, references, bytes, array length & B \\
A3 & Parser Filter & Class blocklist in the parser's own filter & B \\
\midrule
S1 & Execution Limits & Deny process-execution system calls & L \\
S2 & Write Confinement & Read-only root, explicit writable mounts & L \\
S3 & Egress Isolation & Internal network, no external egress & L \\
\bottomrule
\end{tabularx}
\end{table}

Per-layer adapters compile each rule of Table~\ref{tab:catalog-main} into native configuration, reject unsupported controls or scopes, and confirm that the configuration has loaded. A loaded rule counts as acting only when its enforcement event appears during replay. G1 and G2 become NGINX limits and ModSecurity rules~\cite{owaspcrs}, the latter enabled only if they match no training request. A1 and A2 become \textit{jdk.serialFilter} settings~\cite{jep290,oracleSerializationFiltering} and apply only where the vulnerable path uses \textit{ObjectInputStream}. A3 extends the class blocklist of the Jenkins remoting \textit{ClassFilter}, which Jenkins applies to XStream alongside its JEP-200 allowlist. S1-S3 become a seccomp profile denying \textit{execve}~\cite{dockerSeccomp}, a read-only root filesystem with explicit writable volumes, and an internal Compose network. Allowlists, limits, and writable volumes are derived from training workflows only.

\subsection{Running Example}
\label{sec:impl:example}

\begin{figure*}[t]
  \centering
  \includegraphics[width=\textwidth]{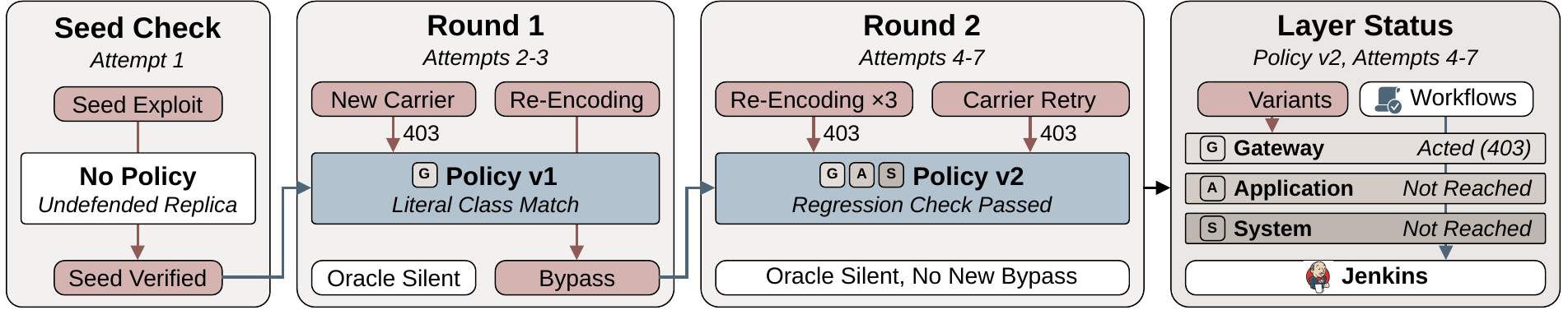}
  \caption{Recorded run on Jenkins CVE-2026-84645. A re-encoded payload bypasses
  the literal gateway rule of v1. The defender revises from the exploited effect into the
  three-layer v2, which stops every later variant at the gateway while benign
  workflows still pass.}
  \Description{Four panels: seed check, round 1 with a gateway-only policy and one
  bypass, round 2 with a three-layer policy blocking all candidates, and the layer
  status showing only the gateway acted while workflows reach Jenkins.}
  \label{fig:running}
\end{figure*}

Figure~\ref{fig:running} traces the J06 run on CVE-2026-84645, a recent XStream
deserialization RCE in Jenkins 2.567~\cite{jenkinsAdvisory202609}. The exploit
posts job-configuration XML from which Jenkins rebuilds a build-trigger object
graph, and the oracle checks a marker file at the sink.

\para{Red-Blue Escalation.} The seed succeeds on the undefended replica, so the
defender installs v1, a gateway rule matching the literal build-trigger class name.
The attacker sees its new carrier rejected and re-encodes the class name as XML
character references. XStream decodes this request to the same object, and the
oracle fires. The defender then revises from the exposed effect rather than from the
bytes of the request. Its v2 adds the class to the Jenkins class filter (A3), which
acts after decoding and so covers every representation of the class, keeps a
decoded-name gateway rule as a cheap pre-filter, and confines writes outside the
Jenkins home as a backstop. v2 is accepted only after it blocks both archived attacks
and passes the benign workflows. The attacker's four further variants all end in HTTP
403, and the run stops by catalog exhaustion. Each attack thus sharpens the next
defense, and each defense narrows what the next attack can use.

\para{Why Baselines Fall Short.} The frozen v2 blocks all four held-out
attacks of J06 and passes all four functionality tests. CRS stops only one
re-encoded variant through literal matching and flags a legitimate
configuration post. Both RASPs preserve functionality but block no attacks,
as their generic hooks do not capture XStream's object reconstruction.
Kintsugi and Tetragon block three attacks at the sink but miss the write
inside Jenkins home, while their process-execution restrictions break a
build workflow. These baselines either miss variants or restrict required
behavior. \sys scopes v2 to the build-trigger payload while preserving legitimate job-management workflows.

\para{Layer Attribution.} Masked replay of the two archived attacks shows that each
is covered by at least two rules, the gateway and application rules for both and the
system rule for the seed, so v2 has no single point of failure. Replaying the
held-out attacks after freezing confirms that the lower layers are not redundant.
With the gateway masked, the application rule alone stops all four and the system
rule two, whereas the gateway alone stops one. The system rule misses the attack
that writes inside the Jenkins home, which builds need, and one variant outside its
narrow workflow-preserving scope.

\section{Evaluation}
\label{sec:eval}

We organize the evaluation around four questions:
\begin{itemize}
\item \textbf{RQ1:} How do frozen defenses compare on held-out attacks and unseen attackers?
\item \textbf{RQ2:} What functionality and online/offline costs accompany their protection?
\item \textbf{RQ3:} How does refinement progress, and what protection remains on paths never rediscovered?
\item \textbf{RQ4:} How do the components, enforcement layers, and budgets affect the outcome?
\end{itemize}

\subsection{Experimental Setup}
\label{sec:eval:setup}

\para{Deployments.} We study sixteen (application, CVE) deployments across six
applications, five on the JVM and one in PHP: Apache OFBiz (4 CVEs), Jenkins (5),
Liferay Portal (2), Metabase (1), Apache ActiveMQ (2), and Snipe-IT (2). Each
reproduces from a fresh snapshot, has a non-destructive consequence oracle and
assertable benign workflows, and has at least one supported control on its exploit
path (Appendix~\ref{app:bench}). One ActiveMQ deployment is reached through native
OpenWire rather than HTTP, so gateway controls do not apply to it.

\para{Attacks and Workflows.} For each deployment an expert constructs five attack
objectives from the same vulnerability, spanning remote code execution, file read,
write, or deletion, request forgery, access-control bypass, privilege escalation,
and application-state abuse~\cite{cwe502,owaspWSTGDeserialization}. One is the
training seed, and the other four are held-out test attacks that differ from it in
impact and, where possible, in gadget family, giving 64 held-out attacks. Each
deployment also has 12 training workflows for the acceptance gate, 4 validation
workflows for baseline tuning, and 4 disjoint functionality tests that assert
response and persistent state. The oracle monitors every consequence reachable on
the path, and only preventive blocks count (Appendix~\ref{app:evaldetails}).

\para{Baselines.} B1 is ModSecurity with OWASP CRS~\cite{owaspcrs}, B2 and B3 are
Datadog RASP~\cite{datadograsp} and OpenRASP~\cite{openrasp}, B4 is
Kintsugi~\cite{peng2026kintsugi}, the closest multi-layer system (\Al+\Sl), and B5 is
Tetragon~\cite{tetragon} with a service-wide policy that denies process execution,
writes outside data directories, and external connections. Tunable baselines receive
the same archive, validation workflows, and tuning allowance
(Appendix~\ref{app:baselines}). B1 does not apply to the OpenWire deployment.

\para{Attackers and Budgets.} Three unseen automated attackers, PentestGPT and CAI on
GPT6 Luna and the CVE-Bench reference agent on Grok 4.7, attempt every deployment,
giving 48 attacker/deployment pairs, and a policy-aware setting reveals the frozen
policy to them. A team-member human expert is reported separately. Synthesis is
capped at 700,000 tokens, 800 replays, 600 attacker queries, 90 minutes, 20 rules,
and 8 rounds, with $k=2$, $\tau_b=100\%$ on the training workflows, and
$\epsilon_l=10\%$. The default backend is DeepSeek V4 Flash.

\para{Metrics.} ASR is the fraction of held-out attacks that still succeed, averaged
over deployments. Mitigation counts deployments with no held-out success and no
synthesis failure, and breach counts attacker/deployment pairs with at least one
success. A failed synthesis is scored with its last accepted configuration, and
blocking and impact limitation are reported separately. BPR is
the fraction of functionality tests passed, FPR the fraction of benign requests
rejected, and $\Delta$P95 the P95 latency increase of the targeted endpoint. No
regression counts deployments where all four tests pass within the latency gate.
Experiments that re-run synthesis are repeated three times and report means with
the largest single-run deviation, and single-run results use the median run.

\begin{table*}[t]
\centering
\caption{Frozen defenses on the 16 deployments. Rates are in percent and Offline in minutes. Main and Switch are ASRs on the 12 deployments with an unused monitored consequence, Aware counts breached attacker pairs with the frozen policy revealed, and Human counts deployments breached by the expert. Other P95 is the non-targeted latency increase.}
\label{tab:main}
\footnotesize\setlength{\tabcolsep}{4.2pt}\tabsetup
\begin{tabular}{@{}l|rrr|rrrr|rrrr|rrr@{}}
\toprule
\tabhead \tabheadfirst{l|}{} & \multicolumn{3}{c|}{Held-out security} & \multicolumn{4}{c|}{Changed attack settings} & \multicolumn{4}{c|}{Functionality and latency} & \multicolumn{3}{>{\columncolor{gray!15}[\tabcolsep][0pt]}c@{}}{Overhead} \\
\cmidrule{2-15}
\tabhead \tabheadfirst{l|}{Method} & ASR & Mitigated & Breached & Main & Switch & Aware & Human & FPR & BPR & $\Delta$P95 & No regr. & Other P95 & RPS & \tabheadlast{r}{Offline} \\
\midrule
ModSecurity + CRS & 55.0 & 0/15 & 32/45 & 45.5 & 59.1{\scriptsize$\,\pm$2.3} & 38/45 & 13/15 & 4.5 & 88.3 & 3.2 & 8/15 & 2.0 & -1.5 & 15 \\
Datadog RASP & 48.4 & 0/16 & 29/48 & 39.6 & 55.6{\scriptsize$\,\pm$1.4} & 34/48 & 12/16 & 0.6 & 96.9 & 9.9 & 8/16 & 5.5 & -4.8 & 18 \\
OpenRASP & 50.0 & 0/16 & 32/48 & 41.7 & 59.7{\scriptsize$\,\pm$1.4} & 37/48 & 13/16 & 0.8 & 96.9 & 10.9 & 7/16 & 6.0 & -5.5 & 18 \\
Kintsugi & 20.3 & 6/16 & 17/48 & 14.6 & 31.9{\scriptsize$\,\pm$2.8} & 23/48 & 9/16 & 0.5 & 95.3 & 9.3 & 12/16 & 1.8 & -2.5 & 12 \\
Tetragon & 18.8 & 8/16 & 18/48 & 12.5 & 36.8{\scriptsize$\,\pm$1.4} & 24/48 & 10/16 & 1.2 & 81.2 & 4.6 & 4/16 & 3.5 & -2.0 & 5 \\
\sys & 6.2 & 13/16 & 5/48 & 6.2 & 9.7{\scriptsize$\,\pm$1.4} & 10/48 & 4/16 & 0.3 & 98.4 & 6.9 & 15/16 & 1.2 & -3.0 & 60 \\
\bottomrule
\end{tabular}
\end{table*}

\subsection{RQ1: Effectiveness}
\label{sec:eval:rq1}

As Table~\ref{tab:main} shows, \sys has 6.2\% held-out ASR, versus 18.8-55.0\% for
the baselines, mitigates 13/16 deployments, and is breached on 5/48
attacker/deployment pairs. On the 15 deployments that B1 covers, \sys's ASR is 6.7\%, but baselines fail in different ways. Specifically, request signatures miss semantic variants,
generic RASP hooks miss deserialization paths outside their sinks, and runtime
containment misses consequences inside permitted resources. Tetragon reaches a
slightly lower ASR than Kintsugi (18.8\% versus 20.3\%) but preserves far fewer
workflows (81.2\% versus 95.3\% BPR), so the baselines do not share a single weakness.

Changed attack settings widen these gaps. Switching to an unused monitored
consequence raises \sys's ASR only from 6.2\% to 9.7\%, whereas Kintsugi rises from
14.6\% to 31.9\% and Tetragon from 12.5\% to 36.8\%, because alternative effects inside
their allowed resources remain reachable. With the frozen policy revealed, the
attackers breach 10/48 pairs for \sys and 23-38 for the baselines. The human expert,
reported separately from the automated attackers, breaches 4/16 deployments for \sys
and 9-13 for the baselines.

The residual attacks come from two synthesis failures and one held-out failure. J11
ends with an archived bypass that no candidate covers within budget, and P02 with a
bypass whose file write overlaps a write that required functionality performs. Both
had accepted a configuration that blocks the seed, and these last accepted policies
remain in every aggregate. J07 passes synthesis but retains one held-out success. Of the 60 prevented \sys attacks,
44 are stopped before or during deserialization and 16 by consequence limitation,
11 of them by denied \textit{execve}, 3 by the read-only root, and 2 by denied egress.
Because G and A act before S and every rule only denies, the first 44 are exactly the
attacks that the G and A rules alone block.
No file-read consequence is credited to limitation. The baselines differ in where
they stop an attack. B1 stops all its blocked attacks before deserialization, while
the RASPs, Kintsugi, and Tetragon stop most of theirs at the sensitive operation.
Neither stopping point establishes that an unmonitored effect is safe.

\finding{Coordinated cross-layer controls cut held-out ASR to 6.2\%, about a third of
the best baseline (18.8\%), and leave only 5/48 attacker/deployment pairs breached.
The advantage persists under consequence switching (9.7\% versus at least 31.9\%) and
policy visibility (10/48 versus at least 23/48 breached pairs), and all residual
attacks trace to two synthesis failures and one held-out failure.}

\subsection{RQ2: Functionality and Overhead}
\label{sec:eval:rq2}

\sys passes 98.4\% of functionality tests at a 6.9\% targeted P95 increase and has
no regression on 15/16 deployments (Table~\ref{tab:main}).
Kintsugi and Datadog RASP also meet the aggregate reference but regress on 4 and 8
deployments, and OpenRASP exceeds the latency gate on average. High functionality
does not imply security, since both RASPs keep 96.9\% BPR at about 50\% ASR. Tetragon
reaches only 81.2\% BPR because some features need subprocesses, writes, or network
access. It rejects only 1.2\% of benign requests, so a low FPR does not establish
functionality preservation, because a request may be admitted while a later
operation or persistent-state assertion fails. \sys adds 1.2\% non-targeted latency
and lowers throughput by 3.0\%. Its 60-minute median offline cost is paid once per
deployment and workflow version, not online. A typical synthesis uses 420,000 tokens and 680 environment replays, 110 of them
counterfactual.

\finding{\sys combines the lowest ASR with the highest BPR (98.4\%) at a 6.9\%
targeted latency increase and regresses on only 1/16 deployments. High BPR alone does
not imply protection, as both RASPs keep 96.9\% BPR yet leave about half of the
held-out attacks successful.}

\begin{figure*}[t]
\centering\includegraphics[width=\textwidth]{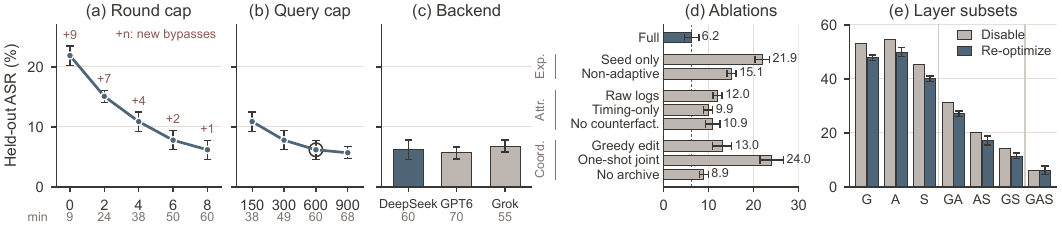}
\caption{Refinement, sensitivity, and ablations (held-out ASR, \%). (a) Round cap,
with new verified bypasses found after each snapshot. (b) Attacker-query cap and
(c) backend, default circled or dark, with median offline minutes below. (d)
Component ablations. (e) Layer subsets, disabled in the frozen policy or
re-optimized. Error bars show the largest single-run deviation over three runs.}
\Description{Five panels of held-out ASR: round cap, query cap, backend, ablation variants, and layer subsets.}
\label{fig:rq34}
\end{figure*}

\subsection{RQ3: Refinement Dynamics}
\label{sec:eval:rq3}

Figure~\ref{fig:rq34}(a) shows diminishing returns. Held-out ASR falls from 21.9\%
after seed-only construction to 6.2\% at the default cap, while new verified bypasses
decline from 9 to 1. In total 14/16 runs stop without a new bypass, 9 by stagnation
and 5 by catalog exhaustion. The other 2/16 exhaust their budget with unresolved
archive attacks, which are the two synthesis failures. Snapshots are scored after
freezing, so held-out outcomes never guide stopping.

To test generalization beyond rediscovery, we split the 64 held-out attacks by
whether the attacker eventually reached their paths, defined on the path keys of
the registered fixtures (Appendix~\ref{app:referee}). The 19 rediscovered attacks
have 1 success (5.3\%) and the 45 never-rediscovered ones 3 (6.7\%), all on the
synthesis failures J11 and P02. On the 14 accepted syntheses, no never-rediscovered
attack succeeds. J07's success is a held-out payload on a rediscovered path whose
archived attack is blocked, consistent with acceptance.

\finding{Refinement shows diminishing returns, as eight rounds cut ASR from 21.9\% to
6.2\% while new bypasses fall from 9 to 1. Protection carries over to paths the
attacker never rediscovered (6.7\% ASR), and none of these attacks succeeds on the
14 accepted syntheses.}

\subsection{RQ4: Ablation and Sensitivity}
\label{sec:eval:ablation}

\para{Components.} Figure~\ref{fig:rq34}(d) isolates exposure, attribution, and
coordination, which address challenges \ding{182}-\ding{184}. Seed-only construction
(21.9\%) and a non-adaptive variant pool (15.1\%) show that adaptive exposure matters
beyond supplying more variants. Without counterfactual replay, ASR rises to 10.9\% and
BPR drops to 94.8\%, reflecting redundant or poorly scoped controls. Of the masked
replays, 71\% end blocked, 18\% at the consequence, 8\% not reached, and 3\% in error,
and 6\% of coverage entries remain unknown, so redundancy is judged only on
conclusive rows. One-shot joint
synthesis reaches 24.0\% and greedy single edits 13.0\%. The gaps among the
attribution, greedy, and no-archive variants are comparable to run-to-run variation
and indicate a consistent direction rather than a strict ranking. Seed-only
synthesis has no synthesis failures, since every seed can be blocked within the
gates, so its higher ASR comes from variants it never saw. The BPR and latency of all
variants appear in Appendix Figure~\ref{fig:app-costs}.

\para{Layers.} Figure~\ref{fig:rq34}(e) shows that every layer matters. Disabling A,
S, or G in the frozen policy raises ASR from 6.2\% to 14.1\%, 31.2\%, and 20.3\%.
Dropping A matters only on Jenkins, where the class filter (A3) carries the type
constraint that the gateway and system rules each cover only in part. Single layers
are much weaker, partly because six deployments have no A control, so A-only ASR is
at least 37.5\%. S-only re-optimization (40.1\%) trails Tetragon because \sys scopes
system rules narrowly to preserve workflows, and on J07-J09 the seeds' state
changes lie inside the writable Jenkins home, where S cannot limit them. \sys's
gateway rules alone (47.9-53.1\%) are close to tuned CRS, leaving 50.0\% open on the
15 HTTP deployments versus 55.0\% for CRS.

\para{Budgets and Backends.} Figure~\ref{fig:rq34}(b-c) shows that ASR stays within
5.7-6.8\% across the three backends, within run-to-run variation. GPT6 Luna also
drives two of the evaluation attackers, so its row is not an independent-attacker
test. The query cap is
split evenly over the rounds, so it changes how many variations each attack phase
may try rather than the number of rounds. Lowering the query
cap to 150 raises ASR to 10.9\%, while raising it to 900 lowers ASR only to 5.7\% at
13\% more offline time. Most runs stop by stagnation or catalog exhaustion before
the overall cap binds, so a larger allowance mainly helps rounds that previously ran
out of queries. Rotating the training seed on six deployments leaves
4.2-8.3\% ASR on their held-out attacks, versus none under the default seed.

\finding{Every component and every layer contributes. The seed-only, non-adaptive,
and one-shot variants cost the most (15.1-24.0\% ASR), and dropping any single layer
raises ASR to 14.1-31.2\%. Results stay within 5.7-6.8\% across backends, and raising
the default 600-query cap to 900 lowers ASR by only 0.5 points.}

\section{Discussion}
\label{sec:discussion}

\subsection{Implications}
\label{sec:discussion:implications}

The evaluation also shows why single-point and generic defenses fall short, which
yields three lessons beyond \sys.

\para{Place the Decisive Constraint Where the Attack Becomes Visible.} A literal
gateway rule misses an equivalent representation, while a type-level constraint after
decoding covers every encoding of the same class, and disabling the application layer
raises ASR from 6.2\% to 14.1\%. A gateway signature is better treated as a cheap
pre-filter that lets downstream controls be scoped narrowly, a principle defenders
can adopt without \sys.

\para{Judge Mitigations on Coverage and Functionality Together.} Frozen system
hardening denies operations that legitimate features need (81.2\% BPR for Tetragon),
whereas RASP keeps 96.9\% of workflows but misses about half of the held-out
attacks. Mitigations should thus be measured on held-out paths, unseen attackers, and
functionality together, since evaluating them on their own seed overstates protection.

\para{Counterfactual Attribution Is Useful on Its Own.} Rule count is a poor proxy for
protection, because an upstream rule hides whether downstream rules hold.
Counterfactual replay needs only replicas and oracles, 94\% of its entries
were conclusive in our runs, and operators can apply it to hand-written policies to
expose redundant rules and single points of failure. These lessons rest on a bounded
benchmark and empirical attackers, whose limits we discuss next.

\subsection{Limitations and Future Work}
\label{sec:discussion:limitations}

\sys preserves only the behavior its benign workflows express, so we use expert-written
workflows with state assertions and unseen tests. Its
protection is empirical rather than proven, and public exploits may appear in model
training data, which independent held-out paths and unseen attackers
partly address. Impact limitation leaves the vulnerable path reachable, so frozen
configurations are temporary measures until a root-cause fix. The benchmark covers
six applications, only one in PHP, so we report per-deployment results rather than
statistical tests. The loop needs only a consequence oracle, a control catalog, and
benign workflows, which makes template injection and request forgery natural next
targets.

\section{Related Work}
\label{sec:related}

\subsection{Deserialization Defenses}
\label{sec:related:deser}

Gadget-chain discovery tools target
PHP~\cite{dahse2014pop,park2022fugio} and Java~\cite{cao2023oddfuzz,chen2023tabby,chen2024jdd,srivastava2023crystallizer},
and empirical studies show that chains are diverse and that their applicability
hinges on classpath and version
prerequisites~\cite{sayar2023deserstudy,kreyssig2024prerequisites,cao2023overriding}.
On the defense side, look-ahead agents restrict the classes a stream may
instantiate~\cite{serialkiller}, runtimes provide class and graph
filters~\cite{jep290}, and
recent systems synthesize them automatically: DeseriGuard derives Java allowlists
from application semantics~\cite{zhang2024deseriguard}, QUACK infers allowed PHP
classes by static duck typing~\cite{david2024quack}, and Gauthier and Bae block
chains at run time~\cite{gauthier2022runtime}. These filters act only at the call
sites and parsers they hook, whereas \sys uses them as one of three layers.

\subsection{Runtime Mitigation}
\label{sec:related:mitigation}

Earlier work blocks known exploit inputs with network and host
filters~\cite{wang2004shield,costa2007bouncer}, disables a
vulnerable path until a patch ships~\cite{huang2016talos,huang2019senx}, or rewrites
code with language models~\cite{nong2025appatch}. System-call
specialization shrinks the attack surface of applications and
containers~\cite{bulekov2021saphire,lei2025rightcalls,ghavamnia2020confine},
while RASP~\cite{datadograsp,openrasp} and eBPF enforcement~\cite{tetragon} apply
generic policies that ignore the specific vulnerability and deployment.
Kintsugi, the closest system, places kernel-enforced policies at application code
locations~\cite{peng2026kintsugi}. These systems fix enforcement points in advance,
whereas \sys selects and scopes controls across layers and revises them against an
adaptive attacker.

\subsection{Cross-Layer Synthesis}
\label{sec:related:crosslayer}

Policy generation for defense in depth~\cite{rubel2005defenseindepth}, firewall
anomaly analysis~\cite{al2004discovery}, and coordinated
front-end filtering~\cite{meng2018rampart} share our premise that layers must be
configured together, but they reason over static rule models and cannot tell whether
a seemingly redundant rule is the only one that stops a variant, which \sys learns
from counterfactual replay. Adversarial testing evades
WAFs~\cite{demetrio2020wafamole,appelt2018waftesting}, and language-model agents
exploit real
vulnerabilities~\cite{deng2024pentestgpt,fang2024oneday,liu2025pacebench,zhu2025c},
but these works measure or break defenses rather than feed variants back into
synthesis. Our loop follows counterexample-guided
synthesis~\cite{solarlezama2006sketching}, adversarial
training~\cite{madry2018towards}, and self-refining
agents~\cite{shinn2023reflexion}, but it produces a deployable
configuration, verifies every counterexample with an independent oracle, and gates
every revision on benign workflows.

\section{Conclusion}
\label{sec:conclusion}

We presented \sys, a red-blue adversarial framework that synthesizes
deployment-specific cross-layer defenses against unsafe deserialization. A black-box
attacker supplies verified exploit variants, and a defender turns them into joint
gateway, application, and system controls through attack-effect summaries,
counterfactual coverage attribution, and regression-gated revision. On sixteen
deployments across six Java and PHP applications, \sys lowers held-out attack success
to 6.2\%, about a third of the best baseline, mitigates 13 of 16 deployments, and is
breached on only 5 of 48 unseen-attacker pairs, while passing 98.4\% of functionality
tests at a 6.9\% targeted latency increase. Its advantage holds under consequence
switching and policy visibility and carries over to exploitation paths the attacker
never rediscovered.

\bibliographystyle{ACM-Reference-Format}
\bibliography{www2027}

\appendix
%% Appendix. Only material that the main text cites and cannot carry itself.

\raggedbottom
\section{Benchmark Details}
\label{app:bench}

Candidates come from the Kintsugi artifact~\cite{peng2026kintsugi},
PACEbench~\cite{liu2025pacebench}, CVE-Bench~\cite{zhu2025c}, and public
advisories. A candidate is kept only if its seed reproduces three times in a row from
a fresh snapshot and it meets the criteria of the evaluation setup.
Table~\ref{tab:benchfull} lists the manifest, and Table~\ref{tab:families} the
enforcement reach and gadget families per application.

\begin{table}[t]
\centering
\caption{Enforcement reach per application. \Al{} covers Java-native paths (A1/A2)
and the Jenkins XStream path (A3); \Wl{} needs an HTTP entry. ActiveMQ's Jolokia CVE
admits \Wl\Sl, its native OpenWire CVE only \Sl.}
\label{tab:families}
\footnotesize\setlength{\tabcolsep}{4pt}\tabsetup
\begin{tabularx}{\columnwidth}{@{}l|l c c >{\raggedright\arraybackslash}X@{}}
\toprule
\tabhead \tabheadfirst{l|}{Application} & Format & \Al{} & Layers & \tabheadlast{l}{Gadget families} \\
\midrule
OFBiz & Java-native & \ding{51} & \Wl\Al\Sl & CommonsBeanutils, ROME, MozillaRhino \\
Jenkins & XStream & \ding{51} & \Wl\Al\Sl & nested PersistenceRoot, OfflineCause, ListView, Agent identity, lazy render \\
Liferay & JSON & \ding{55} & \Wl\Sl & C3P0/ReferenceSerialized, JdbcRowSetImpl \\
Metabase & Java-native & \ding{51} & \Wl\Al\Sl & Java URLDNS, Clojure comparator \\
ActiveMQ & JMX / OpenWire & \ding{55} & \Wl\Sl/\Sl & Jolokia MBean, ClassPathXmlApplicationContext \\
Snipe-IT & PHP object & \ding{55} & \Wl\Sl & Laravel, Monolog, Guzzle chains \\
\bottomrule
\end{tabularx}
\end{table}

\begin{table*}[!tp]
\centering
\caption{Benchmark manifest. Auth indicates whether the attack requires authentication, and Points lists the catalog enforcement points on the seed's exploit path. S is listed only where the consequence involves process execution, a write outside writable volumes, or egress, so it cannot limit the file read of P01 or the state changes of J07-J09 inside the writable Jenkins home. P01 assumes the default APP\_KEY, and J12 a user permitted to run native H2 queries.}
\label{tab:benchfull}
\small\tabsetup
\begin{tabular}{@{}l|llllcc@{}}
\toprule
\tabhead \tabheadfirst{l|}{ID} & CVE & Application (version) & Format & Consequence & Auth & \tabheadlast{c}{Points} \\
\midrule
\multicolumn{7}{c}{\emph{Java}} \\
\midrule
J01 & CVE-2020-9496 & Apache OFBiz 17.12.01 & Java-native & RCE & No & G,A,S \\
J02 & CVE-2021-26295 & Apache OFBiz 17.12.01 & Java-native & RCE & No & G,A,S \\
J03 & CVE-2021-29200 & Apache OFBiz 17.12.01 & Java-native & RCE & No & G,A,S \\
J04 & CVE-2021-30128 & Apache OFBiz 17.12.01 & Java-native & RCE & No & G,A,S \\
J05 & CVE-2026-53435 & Jenkins 2.567 & XStream & RCE & Yes & G,A,S \\
J06 & CVE-2026-84645 & Jenkins 2.567 & XStream & RCE & Yes & G,A,S \\
J07 & CVE-2026-84646 & Jenkins 2.567 & XStream & State change & Yes & G,A \\
J08 & CVE-2026-84650 & Jenkins 2.567 & XStream & Config change & Yes & G,A \\
J09 & CVE-2026-84651 & Jenkins 2.567 & XStream & Config change & Yes & G,A \\
J10 & CVE-2020-7961 & Liferay Portal CE 7.2.0 GA1 & JSON & RCE & No & G,S \\
J11 & CVE-2020-15842 & Liferay Portal CE 7.2.0 GA1 & JSON & RCE & Yes & G,S \\
J12 & CVE-2026-59827 & Metabase OSS 0.61.1.3 & Java-native & RCE & Yes & G,A,S \\
J13 & CVE-2022-41678 & ActiveMQ Classic 5.17.3 & Jolokia JMX & RCE & Yes & G,S \\
J14 & CVE-2023-46604 & ActiveMQ Classic 5.17.3 & OpenWire & RCE & No & S \\
\midrule
\multicolumn{7}{c}{\emph{PHP}} \\
\midrule
P01 & CVE-2024-48987 & Snipe-IT 7.0.9 & PHP object & File read/write & No & G,S \\
P02 & CVE-2025-59713 & Snipe-IT 7.0.9 & PHP object & RCE & Yes & G,S \\
\bottomrule
\end{tabular}
\end{table*}

\section{Evaluation Details}
\label{app:evaldetails}

\para{Attack Objectives.} Objectives follow the deserialization-impact taxonomy
consolidated from MITRE CWE-502~\cite{cwe502} and the OWASP Web Security Testing
Guide~\cite{owaspWSTGDeserialization}. Denial of service is excluded because its
consequence cannot be confirmed by a non-destructive oracle. Where an application
exposes more than one gadget family (Table~\ref{tab:families}), held-out objectives
also differ from the seed in family. Each ActiveMQ CVE exposes a single family, so its
held-out objectives vary the impact and the execution variant instead. Every held-out
attack first reaches its consequence on the undefended replica, and none is added to
the archive.

\para{Oracles.} The oracle monitors process execution, protected-file access,
controlled egress, and an application-state canary where applicable, and an attack
succeeds if any monitored consequence occurs. Access-control bypass, privilege
escalation, and state abuse are confirmed through the canary. No system-layer control
can limit a file read (Table~\ref{tab:aes}), so only G or A rules can prevent such a
consequence. Alerts, HTTP error codes, replay failures, and process termination after
the consequence do not count as prevention.

\para{Attackers and Budgets.} The automated attackers are pinned to the September
2026 model snapshots, and each receives 200 queries, 60 minutes, and 500,000 tokens per
deployment, separate from the synthesis caps. The human expert uses the same external
observation channel, has no access to the archive or held-out suite, and spends eight
hours per deployment. The synthesis query cap is split evenly over the rounds (75 per
round by default), and every query, verification, regression check, and
counterfactual replay consumes environment replays. The consequence-switch suite
covers the 12 deployments that leave a monitored consequence unused, since the
objectives of J07-J09 and P01 already use every monitored consequence on the path.
All runs use one 32-core AMD EPYC server with 128\,GB of memory, and latency is
measured while replaying the training workflows at 50 requests per second.

\para{Metrics and Aggregation.} FPR is per benign request and BPR per functionality
test, so they are not complements, because a request may be admitted while a later
operation fails. Every applicable deployment contributes four attacks, so pooled and
macro ASR coincide. We treat applications as clusters and report per-deployment
counts instead of $p$-values. Repeated experiments pool the 192 held-out attacks of
three runs. Their single-run deviation stays within 2.8 percentage points, and the
full policy leaves 3, 4, and 5 of 64 attacks successful across runs (4.7-7.8\% ASR).

\section{Attack-Effect Summary}
\label{app:aes}

The defender reasons over a reduced view of each attack's trace rather than the raw
system-call stream. \sys keeps only operations linked to the attack request
and projects them onto the capability vocabulary of
Table~\ref{tab:aes}. The vocabulary describes what was observed and is broader
than what the system layer can enforce: the table separately lists which catalog
control, if any, can intercept each capability, so an observed capability does
not imply an available limitation. The decisive
operation, the one the oracle observes, is recorded with its parameters and lineage;
all other operations are collapsed to a count. The oracle, not the summary, decides
whether an attack succeeded.

\begin{table}[t]
\centering
\caption{Capability vocabulary of the attack-effect summary. The last column lists the catalog control that can intercept each capability.}
\label{tab:aes}
\footnotesize\setlength{\tabcolsep}{3pt}\tabsetup
\begin{tabular}{@{}l|p{2.9cm}p{2.1cm}@{}}
\toprule
\tabhead \tabheadfirst{l|}{Capability} & Raw operations & \tabheadlast{p{2.1cm}}{Interceptable by} \\
\midrule
\textit{proc\_exec} & \textit{execve}, \textit{execveat} (also via \textit{posix\_spawn}) & S1 \\
\textit{file\_write} & \textit{openat} (write/create), \textit{rename}, \textit{unlink}, \textit{chmod} & S2, outside writable volumes only \\
\textit{file\_read} & \textit{openat} (read-only) & none \\
\textit{net\_egress} & \textit{connect}, \textit{sendto} (non-loopback) & S3 \\
\textit{other} & (collapsed to a count) & none \\
\bottomrule
\end{tabular}
\end{table}

\begin{figure*}[t]
\centering\includegraphics[width=.93\textwidth]{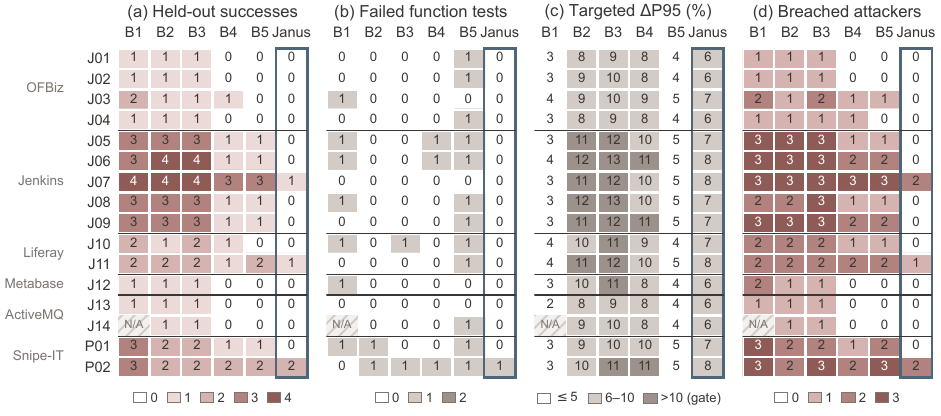}
\caption{Per-deployment results. (a) Held-out successes out of 4 test attacks, all
of which succeed undefended; (b) failed functionality tests out of 4; (c) targeted
P95 increase (\%), with $>$10\% exceeding the latency gate; (d) breached attackers
out of 3. N/A marks B1 on the native OpenWire deployment J14.}
\Description{Four heatmaps of 16 deployments by six methods.}
\label{fig:perdeployment}
\end{figure*}

\begin{figure*}[t]
\centering\includegraphics[width=\textwidth]{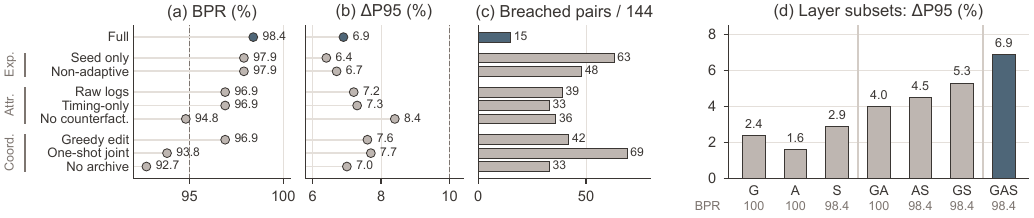}
\caption{Utility and breach counts of the ablations (a-c) and the disabled-layer
subsets (d), complementing the held-out ASR in Figure~\ref{fig:rq34}(d-e).
Dashed lines mark the 95\% BPR and 10\% latency references. Means over three runs;
breaches are summed over 48 attacker/deployment pairs per run.}
\Description{Dot plots of BPR and latency and bars of breached pairs for nine ablation variants, and latency bars for seven layer subsets.}
\label{fig:app-costs}
\end{figure*}

\section{Validation Procedures}
\label{app:referee}

\para{Counterfactual Replay.} \sys replays each archived attack from the same snapshot, masking each rule that blocked it, until the attack reaches its consequence or no rule blocks it. A replay counts as \emph{blocked by $r$} only if
an enforcement event of rule $r$ precedes the point where the consequence would
occur and no consequence is observed; replica or oracle failures are retried a bounded
number of times and never counted as blocks. A replay has four outcomes, and only
two of them determine the remaining entries. If it is blocked by $r$, the entry of
$r$ is obs (first replay) or lat (later replays). If it reaches the consequence,
every still-enabled rule has demonstrably failed to stop $x$ and receives none. If
it ends with neither the consequence nor an enforcement event (\emph{not reached}),
or still fails after retries (\emph{error}), nothing is learned about the enabled
rules, and their entries stay unknown. A none entry thus means that $r$ does not
stop $x$ once the rules that stopped $x$ earlier are disabled. This reading
assumes that disabling a deny-only rule does not reroute $x$ onto a path that $r$
would block.

\para{Attribution Quality.} Each run reports the fraction of masked replays per
outcome, the retries, and the fraction of unknown entries. On 12 reference
configurations with known coverage, such as duplicated rules or rules on disjoint
paths, 141 of 144 entries agree with the reference and the other 3 are unknown. Unknown entries do not affect the measured ASR, but a rule
is declared redundant only from rows without unknown entries.

\para{Path Key.} A verified attack is labeled with its entry point (route and
parameter), the outermost framing of its stream, and a hash of the application
frames between the deserialization call site and the sink. Two attacks with the same
key follow the same exploitation path. Disjointness is checked on the keys of the
catalog fixtures as registered; a permitted variation may produce a different key.
An archived attack whose key equals that of a held-out attack rediscovers that
path. The two attacks may differ in payload, and the
held-out attack is never added to the archive.

\section{Agent Inputs}
\label{app:prompts}

Table~\ref{tab:prompts} lists what each agent sees, never sees, and returns. 
% Prompt templates will be released with the artifact.

\begin{table}[t]
\centering
\caption{Information available to the two agents.}
\label{tab:prompts}
\footnotesize\setlength{\tabcolsep}{4pt}\tabsetup
\begin{tabularx}{\columnwidth}{@{}l|>{\raggedright\arraybackslash}X>{\raggedright\arraybackslash}X@{}}
\toprule
\tabhead \tabheadfirst{l|}{} & \multicolumn{1}{c}{Attacker} & \tabheadlast{c}{Defender} \\
\midrule
Sees & Vulnerability description, reviewed fixtures and prerequisites, own requests and external feedback & Attack-effect summaries, coverage matrix, failed workflow assertions, incumbent rules, catalog \\
\midrule
Never sees & Oracles, defense logs, coverage results, policy text & Held-out attacks, test workflows \\
\midrule
Returns & Fixture choice and parameters in a request schema & Typed edits with evidence and expected effect \\
\bottomrule
\end{tabularx}
\end{table}

\section{Baselines}
\label{app:baselines}

\begin{table}[t]
\centering
\caption{Baseline applicability and configuration.}
\label{tab:baselines}
\footnotesize\setlength{\tabcolsep}{3pt}\tabsetup
\begin{tabularx}{\columnwidth}{@{}c|l c c >{\raggedright\arraybackslash}X@{}}
\toprule
\tabhead \tabheadfirst{c|}{ID} & \multicolumn{1}{c}{Method} & Layer & Coverage & \tabheadlast{c}{Configuration} \\
\midrule
B1 & ModSecurity + CRS~\cite{owaspcrs} & \Wl & 15/16 & Tuned paranoia, exclusions \\
B2 & Datadog RASP~\cite{datadograsp} & \Al & 16/16 & Tuned detection rules, thresholds \\
B3 & OpenRASP~\cite{openrasp} & \Al & 16/16 & Tuned detection rules, thresholds \\
B4 & Kintsugi~\cite{peng2026kintsugi} & \Al+\Sl & 16/16 & Reproduced per target \\
B5 & Tetragon~\cite{tetragon} & \Sl & 16/16 & One fixed service-wide policy \\
\bottomrule
\end{tabularx}
\end{table}

The tunable baselines B1-B3 receive the same 12-trial allowance on the evolved
archive and are selected on validation workflows. Kintsugi inserts trigger markers
into application code, a deployment assumption that configuration-only controls
avoid. The greedy and one-shot ablations share \sys's candidates, model, evidence,
and budget, and greedy search accepts an edit only if it keeps archive ASR and all
thresholds.

\section{Additional Results}
\label{app:results}

All figures and tables are generated from one result file by scripts that also check
denominators, failure inclusion, and the A-only lower bound.

\para{Per-Deployment Results.} Figure~\ref{fig:perdeployment} gives the integer
counts behind every aggregate. \sys's residual failures are confined to J07, J11,
and P02, which account for all four held-out successes and all five breached pairs.
B1-B3 are breached on every deployment they cover and fail most on Jenkins, with 3-4
of 4 held-out attacks succeeding on each of J05-J09. B5 fails a functionality test on
12 of 16 deployments, B2 and B3 exceed the latency gate on 6 and 9, and \sys fails a
test only on P02 and stays within 8\% latency.

\para{Utility of Ablations and Layer Subsets.} Figure~\ref{fig:app-costs} reports
what Figure~\ref{fig:rq34} omits. Every ablation keeps BPR between 92.7\% and 97.9\%,
below the full policy's 98.4\%, and breached pairs follow the held-out ASR ranking,
from 15 of 144 for the full policy to 69 for one-shot synthesis. For the layer
subsets, targeted latency grows nearly additively, and BPR is 100\% whenever S is
disabled because the full policy's only failing test comes from an S rule.

\end{document}